\documentclass[global,twocolumn]{svjour}
\usepackage{latexsym}
\usepackage[dvips]{graphicx}
\usepackage[]{epsfig}
\newcommand*{\degC}{\ensuremath{^\circ}C}

\begin{document}
\title{GaN and InN nanowires grown by MBE: a comparison}
\author{Raffaella Calarco \and Michel Marso
}                     
\offprints{}          
\institute{Institute of Bio- and Nanosystems (IBN1) and cni - Centre of
Nanoelectronic Systems for Information Technology, Research Center J\"ulich,
52425 J\"ulich, Germany}
%
%
\maketitle
\begin{abstract}
Morphological, optical and transport properties of GaN and InN nanowires grown
by molecular beam epitaxy (MBE) have been studied. The differences between the
two materials in respect to growth parameters and optimization procedure was
stressed. The nanowires crystalline quality has been investigated by means of
their optical properties. A comparison of the transport characteristics was
given. For each material a band schema was shown, which takes into account
transport and optical features and is based on Fermi level pinning at the
surface.
\end{abstract}
\section{Introduction}
\label{intro} Intense research concentrating on miniaturization of dimension in
semiconductor devices has been developed in recent years.  This trend is
expected to be limited by fundamental physical constraints. Therefore an
intense effort to search for new manufacturing procedures alternative to
conventional top-down approaches has been strongly motivated.

Self organized bottom-up methods are well suited for the preparation
of structures significantly small (nanometer scale), as required for
device applications based upon quantum effects. Among other
nanostructures semiconductor nanowires have attracted a great deal
of attention. Since the first demonstration in 1964 \cite{Wagner64},
freestanding semiconductor nanowires (NWs) deserved a significant
research attention owing to a quite unique combination of an
intriguing growth mechanism as well as structural and electronic
properties. An advantage of NW heteroepitaxy is that much more
combinations of materials are possible because nanowire synthesis
prevents formation of dislocations originating from lattice
mismatch. The NWs can elastically relax laterally at relatively
short distances from the heterostructure interfaces. Thus greater
lattice mismatch can be accommodated through pseudomorphic growth
without defect introduction when compared to traditional
two-dimensional thin film growth. In addition nanowires can be
fabricated on a wide variety of substrates, including silicon, which
make them suitable for future CMOS integration.

In recent years a tremendous development of new families of
nanodevices utilizing wire materials is emerging
\cite{Cui03,Greytak04,Hahm04,Huang02,McAlpine03,Zhong03a,Duan03,Zhong03,Bjoerk02,Bjoerk02a,Calarco05,Thelander05}.
III-Nitride based nanowires are also investigated as potential
nanoelectronic devices. GaN nanowires with extremely good crystal
quality and strong luminescence efficiency
\cite{Calarco04,Calleja00,Calleja99,Sanchez-Paramo02} have already
been grown by MBE on different substrates. While the nanowire growth
by MBE has already been established, a lot of uncertainty remains on
the mechanisms driving the growth. In this context our investigation
of plasma-assisted MBE (PAMBE) grown GaN nanowires \cite{Meijers06}
demonstrates the possibility to tune the physical properties of
nanowires reaching tapering effect control and wires crystalline
quality improvement. GaN nanowires with a wide range of
heterostructure geometry and composition can also be fabricated with
good reproducibility \cite{Ristic03,Ristic05,Ristic02,Ristic02a}.
The ability to obtain both p and n-type doping of nanowires is
crucial for electron and hole current injections and light emission
by interband transitions. This has been demonstrated for GaN and
high Ga-content InGaN nanowires by Lieber's group
\cite{Zhong03a,Qian04} constructing either complementary crossed NW
p-n structures or core-shell nanowire heterostructures.
Metal-organic chemical vapor deposition has been used. Light
emission but also subwavelength spatial resolution sensors were
obtained with crossed NW structures \cite{Qian04,Hayden06}.
Additionally the observation of ultraviolet-blue laser action in
single monocrystalline GaN nanowires was reported \cite{Johnson02}.

Among III-Nitrides InN exhibits interesting properties such as low toxicity and
high mobility, which make it suitable for new high-performance devices
\cite{Bhuiyan03}. Islands and nanowires have been prepared by different methods
\cite{Dimakis04,Grandal05,Johnson04,Liang02,Norenberg02,Tang04,Zhang05a}. The
formation of InN nanowires by PAMBE growth has been investigated
\cite{Stoica06} to find out optimum growth conditions for obtaining uniform
wires with high crystalline quality.

Even though sophisticated device structures have already been demonstrated
based on nanowires, many fundamental questions regarding their crystalline and
electronic structure, the influence of internal polarization and electric field
on electronic states, the effect of the large surface with respect to the bulk
and size dependent transport phenomena remain open to a large extent. In this
context, previous investigation on GaN nanowires \cite{Calarco05} demonstrates
the effect of surface Fermi-level pinning and its interplay with the nanowire
dimensions on the recombination behavior of electron-hole pairs in
photoconductivity through these nanowires. Particular emphasis has been given
to the investigation of effects due to space charge layers in order to use them
as design parameters for device performance. As an example, the combination of
narrow gap (InN) and wide band gap (GaN) materials in heterostructure wires, is
of special interest because of the interplay between accumulation and depletion
space charge layers.
\section{Experimental details}
\label{sec:1} Nanowires may be formed in a so-called
vapour-liquid-solid growth mode \cite{Givargizov75}. In this mode,
the growth area is limited by the lateral size of a seed. By
absorbing the growth species and binding them into an eutectic
alloy, this seed provides the source atoms to the growth front
underneath. The seed element can appear either as segregation of one
component of the nanowire material, for example Ga for GaN
\cite{Ristic02}, or as artificial Au droplet \cite{Shirai99}
deposited onto a substrate prior to the growth. In case of MBE grown
GaN nanowires has been, far to our knowledge, not proven if the seed
is a liquid Ga droplet or a few atoms small GaN cluster.

GaN and InN nanowires presented in this paper were grown on Si(111) substrates
by PAMBE using a radio frequency plasma source to activate the nitrogen and
standard Knudsen effusion cells for Ga and In. The growth chamber was pumped
down to a base pressure of $5 \times 10^{-11}$ mbar. Nitrogen partial pressure
in the growth chamber was stabilized during the growth at $3 \times 10^{-5}$
mbar. Silicon (111) substrates were cleaned before being loaded into the MBE
system using a standard chemical cleaning procedure and outgassed in the growth
chamber at $925\degC$ for 15 minutes. A low-energy electron diffraction pattern
shows a clear $7 \times 7$ surface reconstruction, typical of the Si(111)
orientation. NWs are fabricated when growth proceeds under nominal N-rich
conditions. The N-rich growth conditions are obtained at a $N_2$ flux of 4.0
sccm and plasma cell forward power of 500 W.

Cathodoluminescence (CL) and SEM cross-sectional images were
performed in a Leo 1550 SEM equipped with a Zeis VIS grating
monochromator, a CCD camera and a He-cooling cryostat ( 6-475 K
range). The electron beam had energies from 2.5 to 25 keV with a
current ranging from 4 pA to 10 nA.

Photoluminescence (PL) spectra were measured using a Fourier
transform spectrometer (BIORAD FTS40) equipped with a cooled Ge
detector and an argon ion laser emitting 50 mW at 488 nm wavelength.

After epitaxial growth, the nanowires are released from the native Si(111)
substrate by exposure to an ultrasonic bath and deposited on a Si(100) host
substrate covered with an insulation layer of 300 nm $SiO_2$.
Ti(10nm)/Au(100nm) contacts patterned by electron beam lithography allow the
electrical and optoelectrical characterisation of the nanowires. Current
voltage characterization carried out with and without UV light by Hg-Xe lamp
via a quartz fiber (approximately 15 $W/cm^{2}$) have been performed.

\section{Results and discussions}
\label{sec:2} III/V ratio and growth temperature are important
parameters for the successful realization of NWs. Growth time has
also an influence, not only on the length of the nanowires, but also
on their morphology and optical properties. All those parameters
have to be properly chosen to optimize and tune the MBE growth of
III-Nitride based NWs.
\subsection{Growth and optical properties}
\label{sec:3} A wide range of growth parameters was investigated to control the
quality of the wires \cite{Meijers06}. For GaN wires III/V ratio and growth
temperature are essential for tuning the nanowires density; a Ga-flux ramp
during the deposition can instead control the tapering or coalescence growth
modes. Using high resolution CL imaging (Figure 1a) we demonstrate that GaN NWs
with higher crystalline quality (reduced yellow and donor acceptor pairs DAP
emissions in respect to band edge emission $D^0X$) are obtained at higher
deposition temperature. A typical morphology of the MBE GaN nanowires is shown
in an oblique cross-sectional SEM micrograph (Figure 1b). Isolated hexagonal
nanowires are quite homogeneously grown on the whole substrate. Almost all the
NWs result vertically aligned along the [0001] direction. The majority of the
nanowires have a length of about 200-500 nm with diameters ranging from 20-150
nm. Some of those wires (not shown here) can also reach a few micrometers in
length. A detailed experimental description of the growth mechanisms related to
growth conditions is described elsewhere \cite{Meijers06}.
\begin{figure}[h]
\includegraphics[width=0.8\columnwidth]{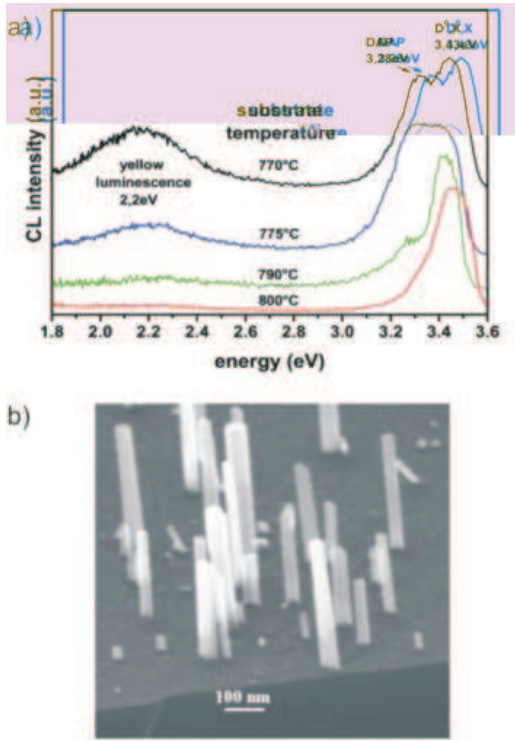}
\vspace{0.1cm}       
\caption{a) CL spectra obtained from GaN samples grown at different substrate
temperature. b) Cross-sectional SEM micrograph of GaN nanowhiskers grown on a
Si(111) substrate.}
\label{fig:1}       
\end{figure}
The MBE growth of InN nanowires takes place in N-rich conditions as for GaN
NWs, but some relevant differences between InN and GaN growth mechanisms have
to be underlined. An important feature is related to the growth temperatures.
Low deposition temperatures in the range $400 - 600 \degC$ are necessary for
InN growth due to a low decomposition temperature of InN
\cite{Dimakis04,Grandal05}, while for GaN the deposition temperature is usually
above $700\degC$. In the case of GaN, the desorption of Ga at high deposition
temperatures is a relevant factor in the growth process. At low temperatures as
in InN MBE growth, the desorption of In can be neglected, while the
decomposition by effusion of nitrogen is a process which strongly affects the
InN growth. This decomposition induces a segregation of In atoms at the free
surface \cite{Dimakis04}. The wire structure thus depends strongly on the
growth temperature. SEM images of three samples grown at different temperatures
are shown in Figure 2. The growth time was 240 min and a beam equivalent
pressure of In of $3.9 \times 10^{-8}$ mbar was used. At a low growth
temperature of $440\degC$, a columnar growth with a relatively high density and
no visible tapering was obtained. At $475\degC$  the wires appear long and
separated from each other showing a good morphology. The increase of the growth
temperature to $525\degC$ results in a low density of columns. A high
non-uniformity in NWs height and shape can be observed. The NWs top region is
flat and shows a well-resolved hexagonal shape the NWs diameter enlarges
towards the top.

\begin{figure}[h]
\includegraphics[width=0.85\columnwidth]{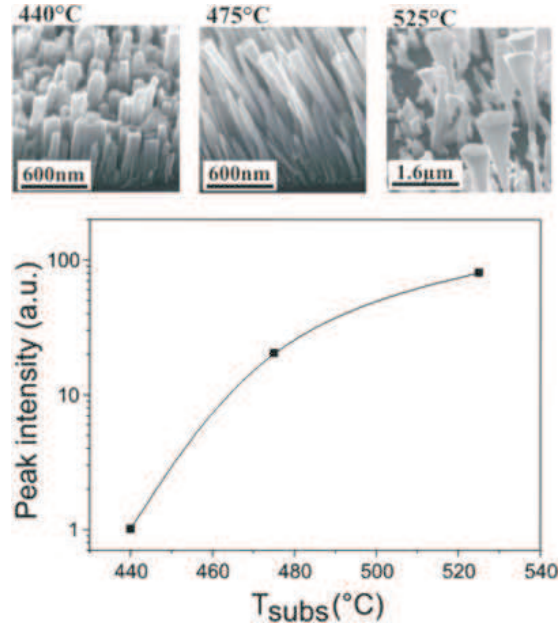}
\vspace{0.1cm}       
\caption{Deposition-temperature ($T_{subs}$) dependence of PL peak intensity.
Cross-sectional SEM micrographs of InN nanowhiskers grown on a Si(111)
substrate at the different substrate temperatures. }
\label{fig:2}       
\end{figure}
\begin{figure}[h!]
\includegraphics[width=0.6\columnwidth]{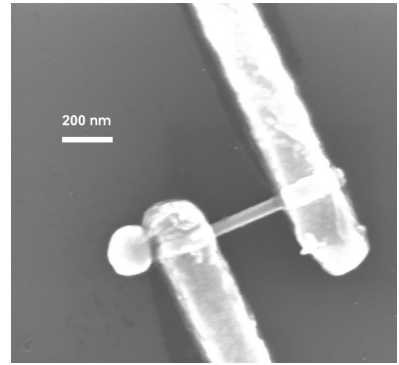}
\vspace{0.5cm}       
\caption{SEM picture of an InN nanowire on Si host substrate, with Ti/Au
contact electrodes. Nanowire diameter: 60nm.}
\label{fig:3}       
\end{figure}
\begin{figure}[h!]
\includegraphics[width=0.85\columnwidth]{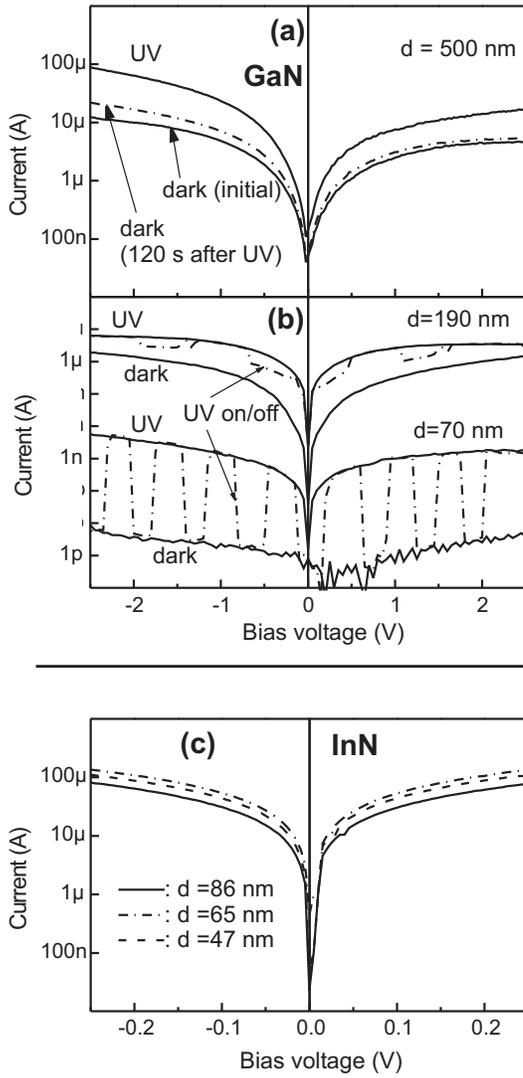}
\vspace{0.5cm}       
\caption{Current-voltage characteristics of nanowires with different diameters.
(a): GaN, 500nm sample, dark and under steady state UV illumination. (b): GaN,
190nm and 70nm samples, dark and under steady state UV illumination, as well as
under periodic UV illumination (dash-dotted). The behavior of the current after
switching off the light (persistent photoconductivity) depends on the diameter.
(c): InN samples: no influence of illumination. Note the different bias voltage
scale of the InN graph.}
\label{fig:4}       
\end{figure}

By increasing the deposition temperature, the PL peak intensity grows more than
two orders of magnitude, as can be seen in Figure 2. The highest PL intensity
accounts to a sample grown at $525\degC$ and is due to well-defined
monocrystalline hexagonal shape of the top part of the wires. Higher deposition
temperatures are usually recommended for high crystalline quality,
corresponding to an intense PL signal. But, at higher growth temperature the
columns are not uniform in diameter and shape. Therefore, the growth of columns
with uniform diameter and high crystalline quality can be obtained only by a
compromise between column uniformity and PL efficiency.

\subsection{Transport properties}
\label{sec:4} Single wire electrical and photoelectrical
investigations have been performed on single nanowire devices with
different diameters \cite{Calarco05,Cavallini06}. Figure 3 shows an
InN nanowire with two ohmic contact electrodes.

Figure 4 exhibits the results of current-voltage and photoconductivity
measurements for nanowires with various diameters d. The GaN devices show a
very strong dependence of the dark current on the diameter. Wires with diameter
above about 100nm have a pronounced persistent photoconductivity, while smaller
devices show a very fast photo response. The InN nanowires, on the other side,
have a much larger conductance than the GaN devices. The dependence of the dark
current on the wire diameter is much weaker than for the GaN wires and the
measured current shows no influence on illumination. These different
(size-dependent) transport properties in GaN and InN originate from the
different band schemata presented in the next paragraph.
\subsection{Band schema model}
\label{sec:5}  The Fermi level pinning at the surface of the GaN NW, about 0.5
to 0.6 eV below the conduction band, creates band bending and a surface
depletion layer (Figure 5a). Since the depletion layer is extended up to 50-100
nm into the bulk, the small diameter wires ($<80$ nm) are expected to be
completely depleted and those with diameters above 100 nm may have a tight open
conducting channel \cite{Calarco05}. This is the reason for the very strong
dependence of the dark current on the wire diameter. In addition a hindered
surface recombination due to the spatial separation of the carriers elucidate
the persistent photocurrent for wires with diameters above 100nm.

For InN, the Fermi level is pinned above the conduction band edge at
the nanowire surface (Figure 5b). Narrow band gap semiconductors
such as InAs and InSb usually show an accumulation layer due to
pinning of the Fermi level above the conduction band edge. The free
electrons move to the surface and form an accumulation layer, in
contrast to the depletion space charge layer in the GaN nanowire.
This results in a very high conductance of the InN nanowires, and
the current is much larger than in the GaN device. This large dark
current also superposes a possible photocurrent.

\begin{figure}[h!]
\includegraphics[width=0.85\columnwidth]{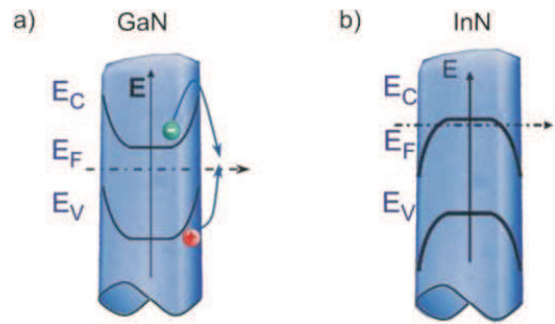}
\vspace{1cm}       
\caption{Single nanowire a) GaN recombination model b) InN band schema.}
\label{fig:5}       
\end{figure}

From PL measurements of InN nanowires discussed in details in our paper
\cite{Stoica06a} a correlation of thermal quenching with PL efficiency was
observed. This effect can be understood by means of a model based on
accumulation layers of a thickness of few nanometers at the wire surface
(Figure 5b). In this surface layer a high-density of electrons is present.
Photoholes are thermally activated over the potential barrier of the downward
bend valance band at the surface and recombine mostly nonradiatively with the
electrons.
\section{Conclusions}
\label{sec:6} The formation of GaN and InN nanowires by MBE growth
are investigated to find out optimum growth conditions for obtaining
uniform wires without coalescence, diameter tapering and/or
enlargement. The NWs optical properties, strongly dependent on the
growth parameters (growth temperature), are used to optimize the
growth process in terms of crystalline quality.

The distinct transport properties of GaN and InN are explained by the different
surface Fermi level pinning leading to surface depletion and accumulation for
GaN and InN respectively.
\section{Acknowledgements}
\label{sec:7} The authors thank K. H. Deussen and H.-P. Bochem for
technical support R. Meijers and T. Richter for measurements and T.
Stoica for valuable discussions.

%
%


\bibliographystyle{nature}
\bibliography{nanocolumns}

\end{document}